\documentclass[fleqn,twoside]{article}
\usepackage{espcrc2}

\usepackage{graphicx}
\usepackage[figuresright]{rotating}

\title{DEPFET Active Pixel Sensors for the Belle II Experiment }

\author{P. Vazquez\address{IGFAE, Dto. F\'isica de Part\'iculas,
        Univ. Santiago de Compostela, Spain, pablo.vazquez@usc.es}
        \thanks{On behalf of the DEPFET collaboration (www.depfet.org)}
        \thanks{Work partially supported by the FPA2008-05979-C04-03 project of the MEYC}}
       
\begin{document}

\begin{abstract}
DEPleted Field Effect Transistor (DEPFET) active pixel detectors combine a first amplification stage with a fully depleted sensor in one single device, resulting in a very good signal-to-noise ratio even for thin sensors. DEPFET pixels are produced in MOS technology with two metal and two poly-silicon layers and have been developed for the use in X–ray imaging and tracking in particle physics experiments. The sensor concept will be presented and all aspects of operation will be detailed with the focus on its application at the upgraded detector Belle II under preparation for the high-luminosity upgrade of the $e^+e^-$ KEKB collider in Japan. The stringent requirements on excellent spatial resolution can be met by cell sizes as small as 25x25 $\mu m^2$ and minimal material budget. The readout ASICs attached to the sensors will be described as well as the module design and the thinning technology employed to reduced the active sensor thickness to as little as 50 $\mu m$.
DEPFET prototype performance at lab and beam tests will be presented, as well as results of irradiations up to 8 MRad. Extensive simulation studies have been made to asses the vertexing and tracking performance of DEPFET-based vertex detector of Belle II. The main performance parameters will be shown, together with an overview of the project status
\end{abstract}

\maketitle

\section{Pixel detector for Belle II experiment}

The Belle experiment, operating at the asymmetric electron positron collider KEKB in Japan, has completed all milestones put forward in 1999. The upgraded machine, SuperKEKB, intended to be operative in 2013, will increase the luminosity by a factor 40 up to 80x$10^{34}cm^{-2}s^{-1}$ by using the nano-beams option. An upgrade of the experiment, Belle II \cite{beltdr}, is also foreseen. In order to cope with the increase of the luminosity the innermost layers of the vertex detector currently made of silicon strips will use pixel technology. 

\begin{figure}[h]
\centering\includegraphics[width=17.5pc]{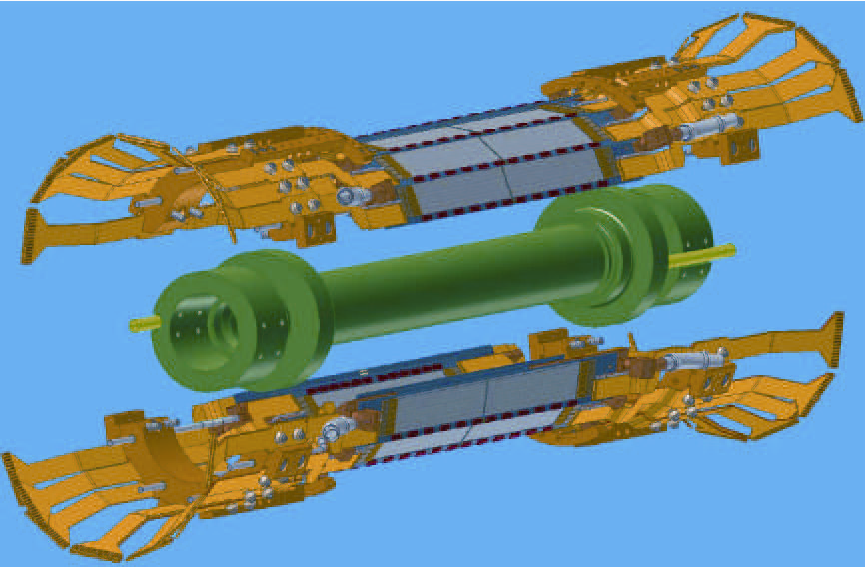}
\caption{The Pixel Detector for Belle II experiment is composed of 2 layers at a radius of 13 and 22 mm mounted on the 10 mm radius beam pipe.}
\label{pxdfig}
\end{figure}

This new pixel detector, PXD (Fig. \ref{pxdfig}), will give a spatial resolution below 10 $\mu$m without sensor active cooling and will contribute to the material budget with less than 0.3 $\%X_0$. The PXD is composed of 8 million 50x75 $\mu m^2$ pixels, placed in 2 layers of 8 + 12 modules each of them read-out from both sides. Every row (sample) is read-out at a rate of 12.5 MHz and the time needed to read a complete frame will not exceed 20 $\mu s$ in order to maintain a low occupancy.

\section{The DEPFET concept and operation}

\begin{figure}[h]
\centering\includegraphics[width=15pc]{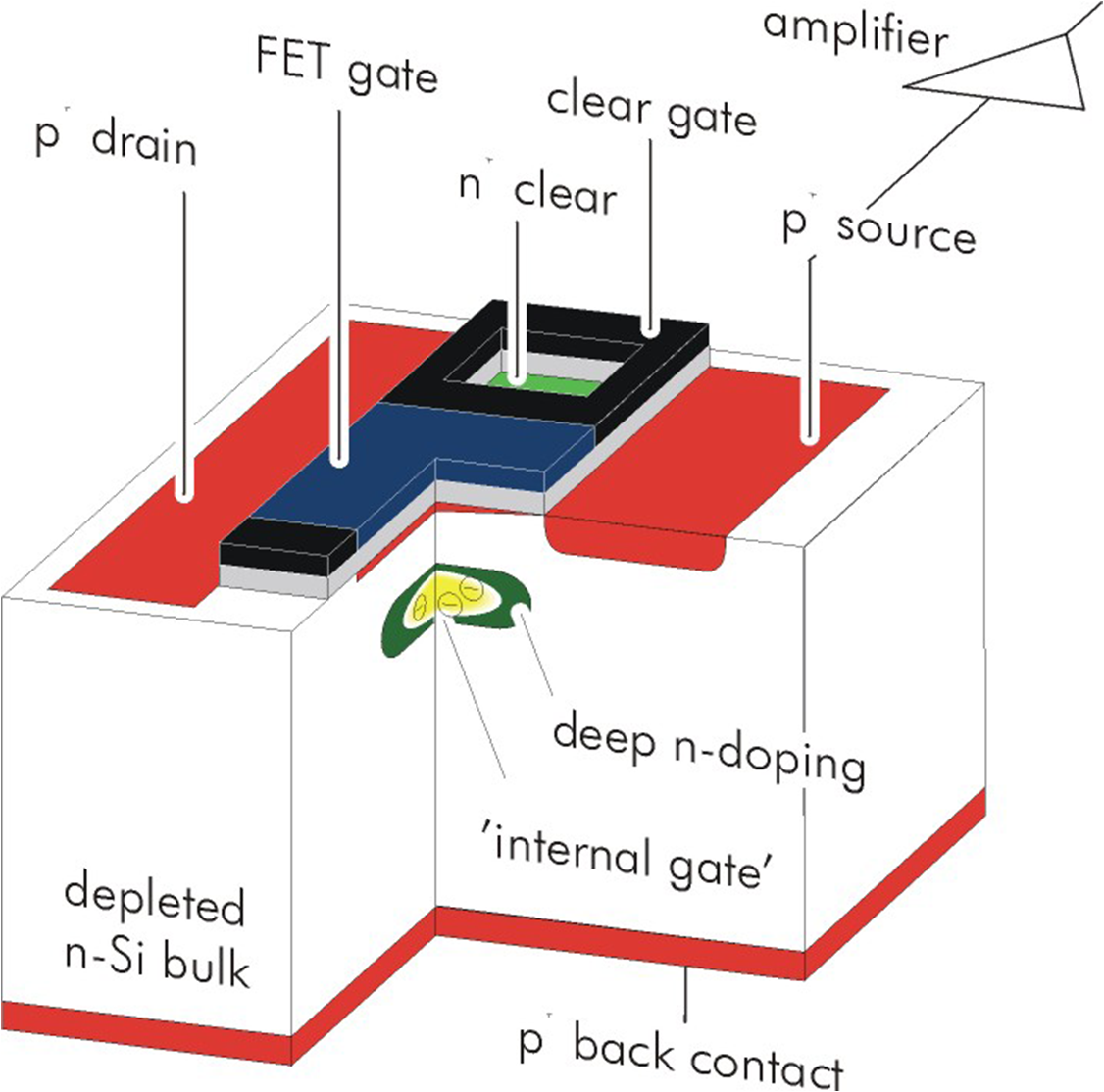}
\caption{The DEPFET detector and amplification structure is based on planar p-channel MOSFET structure on a completely depleted substrate. A deep n-implant acts as an internal gate.}
\label{depfig}
\end{figure}

The  DEPFET concept was first proposed in 1987 \cite{kem87} and further developed in the nineties. The principle of operation is shown in Fig. \ref{depfig}. A MOSFET transistor is integrated onto a detector substrate. By means of sidewards depletion and additional n-implants below the transistor a potential minimum for signal electrons is created underneath the transistor channel. This can be considered as an internal gate of the transistor which modulates the transistor current proportionally to the accumulated charge, this charge can be removed after reading by a clear contact shielded by a p-well.

\begin{figure}[h]
\centering\includegraphics[width=15pc]{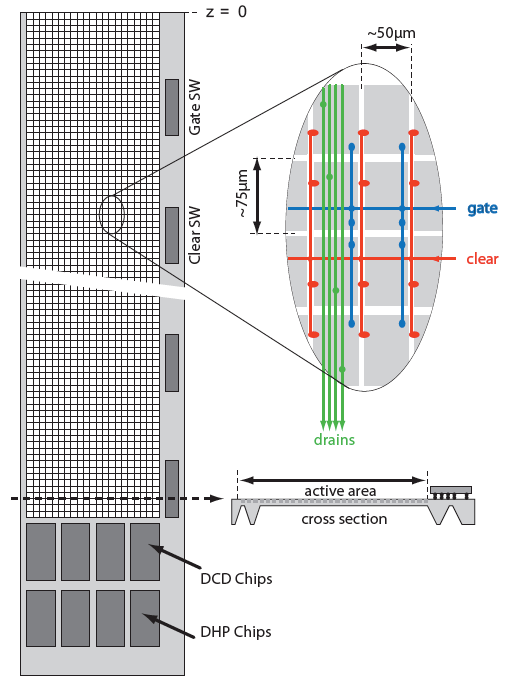}
\caption{Sketch of a PXD half-module for Belle II. All made on silicon, the active is area thinned to 50 $\mu$m and readout ASICs are bump-bonded in the surrounding frame.}
\label{modfig}
\end{figure}

A DEPFET matrix is operated in a rolling shutter read-out mode, all the pixels of a row are accessed at a time by setting a certain potential to their gates. There is no current flow in the non-selected DEPFET rows so that the array consumes very little power. The current at the drain is modulated by the charge collected during the integration time in the internal gate. The pedestal subtraction can be either done by a consecutive Read-Clear-Read sequence (double sampling) or by a faster Read-Clear procedure (single sampling), where digital cached pedestal values are subtracted later. Three different ASICs \cite{kru09} are needed to operate a DEPFET matrix: the Switcher selects, using \lq\lq{high voltage}\rq\rq (10-20V) pulses, the rows to be read/clear; The DCD digitizes drain current signals of pixels from selected row; the DHP reduces the data rate from the DCD and takes care of triggers and synchronization. In order to cope with the 20 $\mu s$ frame read-out time, 4 rows will be addressed simultaneously in Belle II operation (Fig. \ref{modfig}).

\section{Radiation and thinning}

We want to assure PXD components sustain irradiation levels of 10 MRad (TID) and $10^{13}~n/cm^2$ (NIEL) in Belle II. The DEPFET is a MOSFET device and suffers mainly from threshold voltage shifts and also noise increase. Several tests have been done with $\gamma s$, X-rays, protons and neutrons. By reducing the oxide thickness from 200 to 100 nm, we expect to reduce the $\Delta V_{th}$ after 10 MRad from 16 to 3 V which will facilitate the design of the Switcher. The leakage current corresponding to $10^{13}~n/cm^2$ in the matrix correspond to a noise of 200 $e^-$ (or a S/N of 20:1) at a temperature of $27^o$. No problems were seen in DCD2 after being irradiated up to 7.5 MRad or the version 4 of Switcher after 36 MRad. The thinning technology, which allow us to reduce the thickness of the sensing area from 450 to 50 $\mu m$, has been successfully tested on full size electrically active samples.

\section{Lab and beam tests}

Several beam tests \cite{and09} have been carried out in DESY (6 GeV/c) and CERN (120 GeV/c) last years using either Si-strip or DEPFET telescopes. The later composed of 5 planes of 32x24 $\mu m^2$ pixel size covering an area of 2x6 $mm^2$. Different geometries and clear technologies were tested in the beam, giving internal amplification $g_q=\frac{d I_d}{dQ_{int}}$ values between 360-650 pA/e and resolutions down to 1 $\mu m$. Calibration and noise measurements were done in the lab using radioactive sources as $^{109}$Cd or $^{55}Fe$. Laser tests were done to study inter-pixel charge sharing or matrix inhomogeneities. Most of the tests have been done in matrices intended for the ILC.

\begin{figure}[h]
\centering\includegraphics[width=17pc]{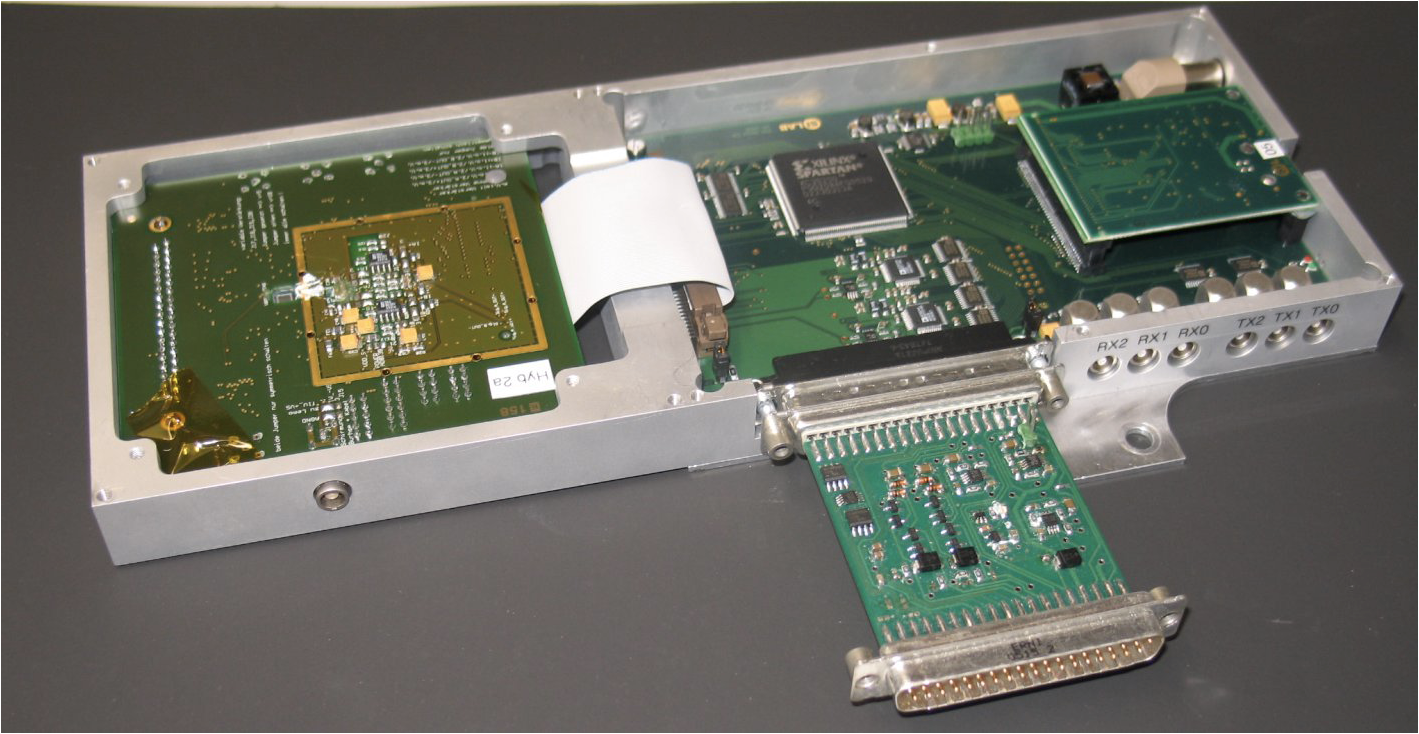}
\caption{DEPFET test module being used in lab and beam tests.}
\label{testfig}
\end{figure}

\section{Simulations}

Predictions from different montecarlo are not consistent in the particle rate at the PXD but the worse scenario does not suggest an occupancy above 0.5\%. Several changes in the baseline scenario were compared using the z-vertex resolution of $J/\Psi$. As a result of these simulations we have seen that it is slightly better to change from 14 to 13 mm the inner layer radius, or that it doesn't matter if the sensor thickness changes from 50 to 75 $\mu m$ as the multiple scattering increase but the fraction of single hits decrease. Simulations also show that the impact parameter resolution of Belle II is twice as good as that of Belle.

\section{Summary}

The DEPFET technology has been presented. The module design for building a pixel detector for Belle II experiment has been described. Prototype performance, irradiation, beam and lab tests as we as some results of simulations were shown. DEPFET matrices and ASICs are approaching their final versions, all of them have to be tested in common with service systems before real detector is going to be produce and install in Japan in 2013.

\end{document}